\documentclass[10pt,journal,twocolumn,twoside]{IEEEtran}
\usepackage{amsmath,amsfonts}
\usepackage{algorithmic}
\usepackage{algorithm}
\usepackage{array}
\usepackage[caption=false,font=normalsize,labelfont=sf,textfont=sf]{subfig}
\usepackage{textcomp}
\usepackage{stfloats}
\usepackage{url}
\usepackage{verbatim}
\usepackage{graphicx}
\usepackage{cite}
\usepackage{color}
\begin{document}

\title{Channel Sensing for Holographic Interference Surfaces based on the Principle of Interferometry}

\author{Jindiao Huang, Yuyao Wu, Haifan Yin,~\IEEEmembership{Member,~IEEE,} Yuhao Zhang and Ruikun Zhang
\thanks{J. Huang, Y. Wu, H. Yin, Y. Zhang and R. Zhang are with Huazhong University of Science and Technology, 430074 Wuhan, China (e-mail: jindiaohuang@hust.edu.cn, yuyaowu@hust.edu.cn, yin@hust.edu.cn, yuhaozhang@hust.edu.cn, zhangrk@hust.edu.cn).}
\thanks{The corresponding author is Haifan Yin.}
\thanks{This work was supported by the National Natural Science Foundation of China under Grant 62071191.}}



\maketitle
\begin{abstract}
The Holographic Interference Surface (HIS) provides a new paradigm for building a more cost-effective wireless communication architecture. In this paper, we derive the principles of holographic interference theory for electromagnetic wave reception and transmission, whereby the optical holography is extended to communication holography and a channel sensing architecture for holographic interference surfaces is established. Unlike the traditional pilot-based channel estimation approaches, the proposed architecture circumvents the complicated processes like filtering, analog to digital conversion (ADC), down conversion. Instead, it relies on interfering the object waves with a pre-designed reference wave, and therefore reduces the hardware complexity and requires less time-frequency resources for channel estimation. To address the self-interference problem in the holographic recording process, we propose a phase shifting-based interference suppression (PSIS) method according to the structural characteristics of communication hologram and interference composition. We then propose a Prony-based multi-user channel segmentation (PMCS) algorithm to acquire the channel state information (CSI). Our theoretical analysis shows that the estimation error of the PMCS algorithm converges to zero when the number of HIS units is large enough. Simulation results show that under the holographic  architecture, our proposed algorithm can accurately estimate the CSI in multi-user scenarios.
\end{abstract}


\section{Introduction}
\IEEEPARstart{T}{he}
sixth generation (6G) wireless communication network needs to meet the requirements on high speed, low latency, and full coverage \cite{ref1}. 6G is expected to build a ubiquitous ultra-broadband green network with ultra-high speed, ultra-high data density, and ultra-low latency. To meet the above requirements, mobile communication systems need to further improve spectrum efficiency and network capacity to achieve terabyte-level distributed computing, 100 Gbit/s/m$^{2}$ or 1-10 Tbit/s/m$^{3}$ ultra-high data density, and ultra-low latency.

In recent years, wireless communication research has shown great enthusiasm for massive multi-input multi-output (MIMO) systems, hoping to use this technology to greatly improve the spectrum efficiency and network capacity of communication systems. Although large-scale phased arrays with hundreds or more antennas have the potential to achieve these goals, they are difficult to deploy effectively in practical applications due to their strong dependence on power amplifiers with enormous power consumption and high-cost phase shifters in high-frequency bands \cite{ref3}. Therefore, in order to meet the demand for massive data transmission in future communication networks, a more economical and efficient communication architecture is needed.

Benefiting from the development of holographic interferometry and metamaterial antennas, holographic communication is expected to provide an innovative communication architecture and support the development of 6G communications \cite{ref2}, \cite{ref30}, \cite{ref34}. 
In this paper, we introduce the Holographic Interference Surface (HIS). The HIS, composed of multiple metamaterial radiation units, detects or estimates wireless signals based on the interferometric superposition of the object wave transmitted by the user equipment (UE) and the reference wave generated by the surface. 
The three-dimensional electromagnetic space is then perceived and reconstructed efficiently by HIS based on the holograms. With its excellent radio frequency (RF) spatial awareness and spatial wave field synthesis capability, the HIS can achieve precise sensing, accurate regulation, and intelligent optimization of the electromagnetic environment, which is likely to provide a revolutionary communication architecture for building a more cost-effective system.
\IEEEpubidadjcol

Holography, a process that forms images based on reconstructed wavefronts, was presented by D. Garbor in 1948 \cite{ref27}, \cite{ref28}. Since holography requires a powerful source of coherent light, the development of the laser greatly facilitated the improvement of holography. Therefore, holography has been widely applied in optics. A multiplexing method using multiple carrier frequencies to form a plurality of reference waves for optical hologram recording has been proposed in \cite{ref7}. The work of \cite{ref29} realizes space-based imaging interferometers based on holography, which avoids multi-spectral detection of optics. Although holography has been extensively researched in optics, the principle of optical holography is not directly available for communication. Doppler and multipath effects have to be considered in communication scenarios, which makes the spectrum composition in communication usually far more complex than in optics. Besides, restricted by the array size, the spatial information available in communication holography is generally much less than the optical, which requires novel algorithms to process holograms based on channel characteristics and time samples.

At present, the research on holographic communication systems focuses on antenna structure design, communication system modeling, and signal transmission. A new holographic antenna design is proposed in \cite{ref9}, which utilizes the polarization of metamaterials to control the antenna state. In \cite{ref3}, the hardware structure and radiation principles of reconfigurable holographic surfaces (RHS) are proposed, which are only available for transmission. 
The authors of \cite{ref10} present a holographic MIMO (HMIMO) communication system enabled by stacked metasurface layers, which is capable of wave-based precoding and combining.
In \cite{ref11}, a channel model for HMIMO arrays based on Fourier transform and classical linear system models is constructed. 
The deterministic and stochastic channel modeling methods for HMIMO communication and several single-user channel estimation approaches are summarized in \cite{ref32}.
The work of \cite{ref8} proposes a holographic MIMO interferometric recording method for multipath scenarios, which focuses on the location of moving objects. To present a holographic radar system, the work of \cite{ref12} generalized the synthetic aperture radar (SAR) framework based on the holographic interference principle to present a holographic SAR system and model. A joint sum-rate maximization algorithm for reconfigurable holographic surfaces-based hybrid beamforming is developed in \cite{ref4}. 
Based on the channel model in \cite{ref32}, the authors of \cite{ref33} discuss the ergodic capacity of a point-to-point HMIMO system and holographic beamforming schemes for four typical HMIMO scenarios.
The work of \cite{ref31} proposed a sensing reconfigurable intelligent surface (RIS)-based channel estimation approach, which estimates the CSI from the interference fringe based on the von Mises distributions.
However, most research only demonstrates the feasibility of holography for electromagnetic wave recording and transmission. Few works have studied the principles of holography for signal receiving and channel sensing in wireless communication.

In this paper, we extend the holographic interference principle in optics to communication and construct the basic architecture of the holographic communication system accordingly. In order to mitigate the influence of the interference components appearing in the electromagnetic interference recording process, we analyze in detail the structural characteristics of the communication hologram interference composition and propose a phase shifting-based interference suppression (PSIS) method. Since the holographic communication architecture adopts a new signal transceiving pattern, a new channel estimation algorithm needs to be constructed to support the subsequent processing such as beamforming. Accordingly, we propose a multi-user channel segmentation algorithm based on extended Prony's method, and a theoretical analysis of its asymptotic performance is made. Our proposed channel segmentation algorithm exploits the spatial structure properties of uniform planar array (UPA) channels to accurately estimate the channel state information (CSI) for each user.

The main contributions of our work are as follows:
\begin{itemize}
\item{We propose a channel sensing architecture of holographic interference surfaces. The principles of holographic interference theory for electromagnetic wave reception and transmission are derived, based on which the optical holography is extended to communication holography.}
\item{We present a phase shifting-based interference suppression method to reduce the self-interference in the holographic recording process, which is based on the characteristics of communication hologram structure and interference composition.}
\item{We propose a multi-user channel segmentation algorithm based on extended Prony's method for HIS, which accurately estimates the CSI in multi-user scenarios by utilizing the spatial characteristics of the UPA channel.}
\item{We analyze the asymptotic performance of our channel segmentation algorithm and prove that the estimation error of our channel segmentation algorithm converges to zero as the number of HIS units increases even with the presence of hologram noise.}
\end{itemize}

The rest of this paper is organized as follows. In Section \uppercase\expandafter{\romannumeral2}, the UPA channel model is introduced \cite{ref6}. In Section \uppercase\expandafter{\romannumeral3} we introduce the basic principles of environment sensing and wave field formation for holographic communication system. In Section \uppercase\expandafter{\romannumeral4}, the hologram interference structure and perturbation removal scheme are proposed. The multi-user channel segmentation algorithm and its performance analysis are described in Section \uppercase\expandafter{\romannumeral5}. Section \uppercase\expandafter{\romannumeral6} contains the simulation results and the final conclusions are drawn in Section \uppercase\expandafter{\romannumeral7}.

\textit{Notation:} Matrices and vectors are denoted by boldface letters. For a matrix $\mathbf{X}$, its conjugate, transpose and conjugate transpose are denoted by $(\mathbf{X})^{*}$, $(\mathbf{X})^{T}$ and $(\mathbf{X})^{H}$ respectively. $(\mathbf{X})^{\dagger}$ stands for the Moore-Penrose pseudoinverse of $\mathbf{X}$. The $ \ell_2 $ norm of a vector or the spectral norm of a matrix is denoted by the same symbol $\|\cdot\|_2$. $\mathbb{E}\{\cdot\}$ stands for the expectation. The Kronecker product of $\mathbf{X}$ and $\mathbf{Y}$ is denoted by $\mathbf{X}\otimes\mathbf{Y}$. diag$\{\mathbf{a_1},\cdots,\mathbf{a_N}\}$ is a diagonal matrix with $\mathbf{a_1},\cdots,\mathbf{a_N}$ at the main diagonal. 
$\mathbb{C}$ is the set of complex number. $\mathbb{N}$ is the set of non-negative integers. $|\cdot|$ and phase($\cdot$) stand for the amplitude and the phase of a complex number respectively. $\mathcal{CN}(\mu,\sigma^2)$ represents the complex univariate Gaussian distribution with the mean $\mu$ and variance $\sigma^2$.

\section{System Models}
We consider $N$ UEs in a certain cell. The base station (BS) is equipped with the HIS composed of $N_v$ rows and $N_h$ columns of radiation units. Thus the number of units at the HIS is $N_t = N_vN_h$. The number of antennas at the UE is restricted to 1 for notational simplicity.
The signal bandwidth consists of $N_f$ subcarriers with spacing $\Delta f$. The channel is composed of $P$ multipaths, with each path having a certain Doppler, angle, delay, and complex amplitude.

\begin{figure}[!t]
\centering
\includegraphics[width=3.5in]{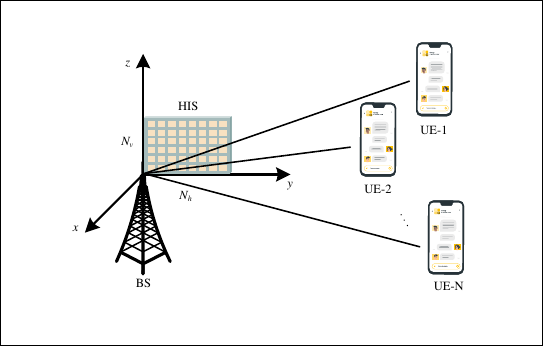}
\caption{The channel between the BS and the UEs.}
\label{fig_0}
\end{figure}

We construct a coordinate system as shown in Fig. \ref{fig_0}. Without loss of generality, the origin is set at the first radiation unit located at the lower left corner of the HIS. The radiation unit index starts from the lower left corner of the surface and increases along the Z-axis until the top row, then continues with the second column, etc. A 3-D steering vector of a certain path with azimuth departure angle $\phi$ and elevation departure angle $\theta$ is defined as
\begin{equation}
    \mathbf{a}(\theta,\phi)=\mathbf{a}_h(\theta,\phi)\otimes\mathbf{a}_v(\theta) \in \mathbb{C}^{N_t},
\end{equation}
where
\begin{multline}
        \mathbf{a}_h(\theta,\phi)=\\
        \begin{bmatrix} 1&e^{2\pi\frac{D_h \sin(\theta)\sin(\phi)}{\lambda_0}}&\cdots&e^{2\pi\frac{(N_h-1)D_h \sin(\theta)\sin(\phi)}{\lambda_0}} \end{bmatrix}^T \in \mathbb{C}^{N_h},
\end{multline}
and
\begin{equation}
    \mathbf{a}_v(\theta)=\begin{bmatrix} 1&e^{2\pi\frac{D_v \cos(\theta)}{\lambda_0}}&\cdots&e^{2\pi\frac{(N_v-1)D_v \cos(\theta)}{\lambda_0}} \end{bmatrix}^T \in \mathbb{C}^{N_v},
\end{equation}
where $D_h$ and $D_v$ are the horizontal and vertical unit spacing of the surface respectively. $\lambda_0$ is the wavelength of the center frequency.

The channel between all radiation units and the antenna of the $u$-th UE at time $t$ and frequency $f$ is denoted by $\mathbf{h}_u(f,t)\in\mathbb{C}^{1\times N_t}$. The channels at all $N_f$ subcarriers can be written in a matrix form $\mathbf{H}_u(t) \in \mathbb{C}^{N_t \times N_f}$:
\begin{equation}
    \mathbf{H}_u(t)=\begin{bmatrix} \mathbf{h}_u^T(f_1,t)&\mathbf{h}_u^T(f_2,t)&\cdots&\mathbf{h}_u^T(f_{N_f},t) \end{bmatrix},
\end{equation}
with $f_i$ being the frequency of the $i$-th ($1\leq i \leq N_f$) subcarrier. According to \cite{ref6}, we may further write
\begin{equation}
\label{eq5}
    \mathbf{H}_u(t)=\mathbf{A}\mathbf{C}_u(t)\mathbf{B},
\end{equation}
where $\mathbf{A}\in\mathbb{C}^{N_t\times P}$ is given by
\begin{equation}
    \mathbf{A}=\begin{bmatrix}\mathbf{a}(\theta_{\text{1,ZOD}},\phi_{\text{1,AOD}})&\cdots&\mathbf{a}(\theta_{\text{P,ZOD}},\phi_{\text{P,AOD}})\end{bmatrix},
\end{equation}
with $\theta_{\text{p,ZOD}}\in [0,\pi]$ and $\phi_{\text{p,AOD}} \in (-\pi,\pi]$ denoting the elevation departure angle and elevation arrival angle of the $p$-th path respectively.
\begin{equation}
    \mathbf{B}=\begin{bmatrix}\mathbf{b}(\tau_1)&\mathbf{b}(\tau_2)&\cdots&\mathbf{b}(\tau_P)\end{bmatrix}^T \in \mathbb{C}^{P \times N_f},
\end{equation}
where $\mathbf{b}(\tau_p)$, ($p=1,\cdots,P$) stands for the delay response vector, which is defined as
\begin{equation}
    \mathbf{b}(\tau_p)=\begin{bmatrix}e^{-j2\pi f_1 \tau_p}&e^{-j2\pi f_2 \tau_p}&\cdots&e^{-j2\pi f_{N_f} \tau_p}\end{bmatrix}^T.
\end{equation}

$\mathbf{C}_u(t)=\text{diag}\{c_{u,1}(t),\cdots,c_{u,P}(t)\}\in\mathbb{C}^{P\times P}$ is a diagonal matrix. Its $p$-th ($p=1,\cdots,P$) diagonal entry is given by
\begin{equation}
    c_{u,p}(t)=\beta_p e^{\frac{j2\pi\hat{r}_{rx,p}^T\overline{d}_{rx,u}}{\lambda_0}}e^{j\omega_pt}.
\end{equation}
$\beta_p$ and $\tau_p$ are the complex amplitude and the delay of the $p$-th path respectively. $\hat{r}_{rx,p}$ is the spherical unit vector with azimuth arrival angle $\phi_{\text{p,AOA}} \in (-\pi,\pi]$ and elevation arrival angle $\theta_{\text{p,ZOA}} \in [0,\pi]$:
\begin{equation}
    \hat{r}_{rx,p}=\begin{bmatrix}
        \sin{\theta_{\text{p,ZOA}}\cos{\phi_{\text{p,AOA}}}}\\
        \sin{\theta_{\text{p,ZOA}}\sin{\phi_{\text{p,AOA}}}}\\ 
        \cos{\theta_{\text{p,ZOD}}}
    \end{bmatrix}.
\end{equation}
$\overline{d}_{rx,u}$ is the location vector of $u$-th UE antenna. $e^{j\omega_p t}$ is the Doppler of the $p$-th path, where $t$ denotes time. $\omega_p$ is given by $\omega_p=\hat{r}_{rx,p}^T\overline{v}/\lambda_0$, where $\overline{v}$ is the UE velocity vector. The velocity vector $\overline{v}$ is given by
\begin{equation}
    \overline{v}=v\begin{bmatrix}\sin{\theta_v}\cos{\phi_v}&\sin{\theta_v}\sin{\phi_v}&\cos{\theta_v}\end{bmatrix}^T,
\end{equation}
 where  $\theta_v$, $\phi_v$, and $v$ are the travel elevation angle, travel azimuth angle and speed of the UE respectively. Therefore, the signal arriving at the HIS is written as
 \begin{equation}
     \mathbf{y}_u = \mathbf{H}_u(t)\mathbf{x}_u + w,
 \end{equation}
 where $\mathbf{x}_u$ denotes the signal transmitted by the $u$-th UE and $w \sim \mathcal{CN}(0,\sigma_w^2)$ denotes the additive white Gaussian noise.

The HIS detects or estimates wireless signals based on the interferometric superposition of the object wave transmitted by UE and the reference wave generated by the surface. The elements of the HIS utilize the same reference wave for interference, and then the intensity of the superimposed signal is recorded by intensity detectors. The interference fringe of RF signal and reference wave contains the phase difference between the two waves, which provides an approach to receive RF signals by only dealing with the envelop of received signal.

\section{Principle of Holographic Communication}
\begin{figure}[!t]
\centering
\includegraphics[width=3in]{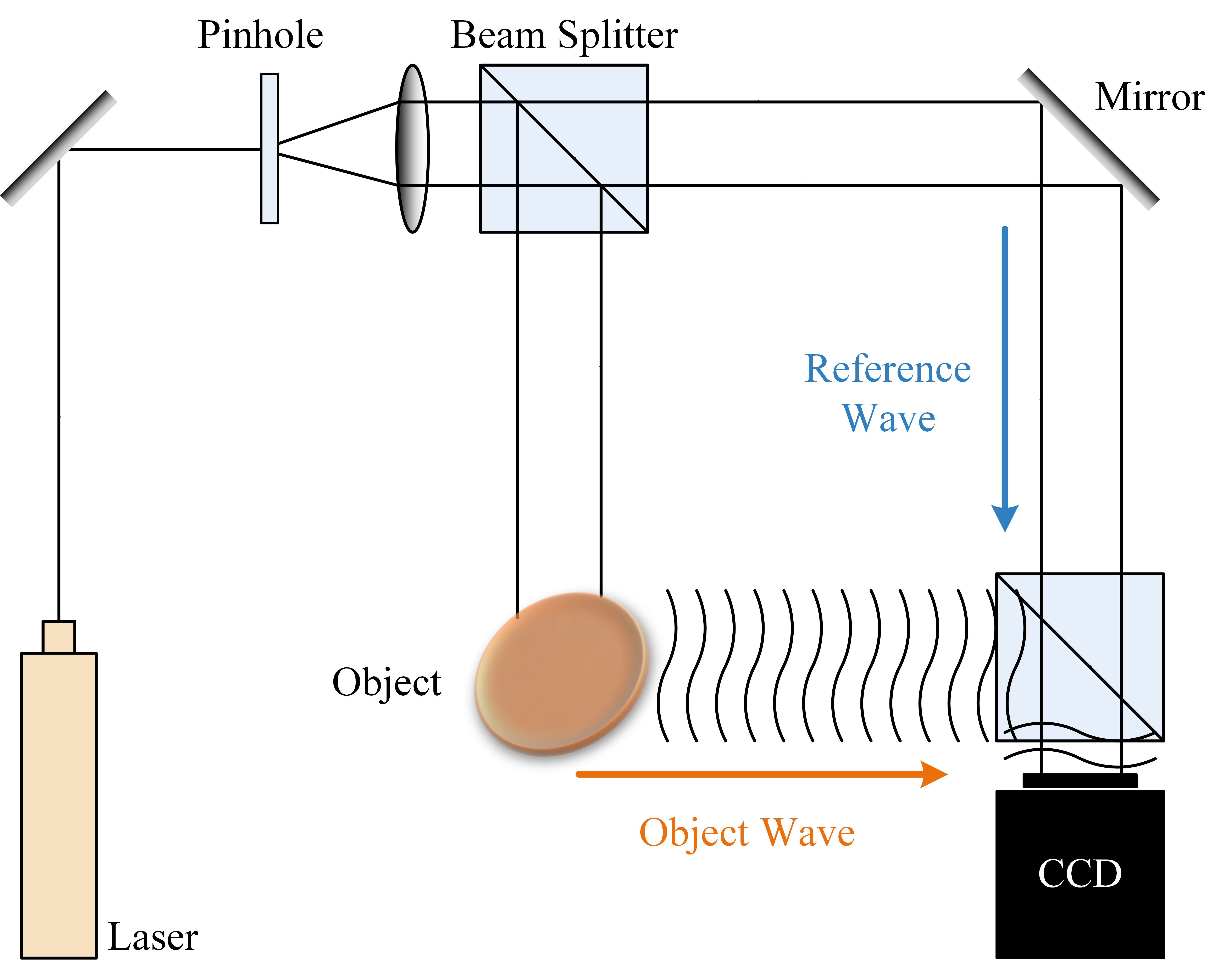}
\caption{Interferometric recording in optical holography.}
\label{fig_1}
\end{figure}

Holography, initially applied in optics, is an imaging technique that utilizes the interference superposition and diffraction propagation of light to achieve high-precision recording and reproduction of three-dimensional objects. The holographic imaging process in optics consists of two basic steps: interferometric recording and diffraction reproduction as shown in Fig. \ref{fig_1}. The interferometric recording uses the interference superposition of the object wave and the reference wave to generate an optical intensity distribution with special interference fringes. The intensity distribution is then recorded as a hologram, which is available for complete recovery of the object wave. Diffraction reproduction is the process of reproducing a specific three-dimensional image using a reference wave to irradiate the hologram. Considering the similarity between optic and electromagnetic waves, optical holography has the potential to be extended to communication, which means that signal can be received and transmitted based on its intensity distribution.

The mathematical expression for electromagnetic wave reception in the holographic communication architecture is derived as follows. When two beams of electromagnetic wave are interferometric superposition in space, a special interference pattern, namely the hologram, is generated,  which records the amplitude and the phase difference between the two waves. The HIS system uses the interference superposition of object wave and reference wave to generate hologram and extrapolate object wave information according to the hologram and reference wave to realize environment sensing.

We adopt the Cartesian coordinate system and place the HIS system composed of $N_v \times N_h$ radiating units on the $zoy$ plane with the $x$-axis perpendicular to the HIS plane. At the $(p,s)$-th unit, the object wave in the environment and the reference wave generated by the unit are given by:
\begin{align}
E_{\text{o}}(\mathbf{r}_{p,s},t) & = A_{\text{o}} \cdot e^{j (\mathbf{k}_{o} \cdot \mathbf{r}_{p,s} + \omega_ot)}, \\
E_{\text{r}}(\mathbf{r}_{p,s},t) & = A_{\text{r}} \cdot e^{j (\mathbf{k}_{\text{r}} \cdot \mathbf{r}_{p,s} + \omega_rt)},
\end{align}
where $A_{\text{o}}$ and $A_{\text{r}}$ are the amplitude of the object wave and the reference wave respectively. $\mathbf{k}_{\text{o}}$ is the directional propagation vector in free space, $\mathbf{k}_{\text{r}}$ is the propagation vector of the reference wave. $\mathbf{r}_{p,s}$ is the position vector of the $(p,s)$-th unit. The interference result of the reference and object waves and its power, namely the hologram, is expressed as:
\begin{align}
\label{eq15}
I(\mathbf{r}_{p,s},t) = &|E_{i}(\mathbf{r}_{p,s},t)|^2 = |E_o(\mathbf{r}_{p,s},t) + E_r(\mathbf{r}_{p,s},t)|^2 \\
  = &|E_o(\mathbf{r}_{p,s},t)|^2 + |E_r(\mathbf{r}_{p,s},t)|^2\\
   &+ E_o(\mathbf{r}_{p,s},t) \cdot E_r^*(\mathbf{r}_{p,s},t)\\
   &+ E_o^*(\mathbf{r}_{p,s},t) \cdot E_r(\mathbf{r}_{p,s},t).
\end{align}

It can be seen from Eq. (\ref{eq15}) that the hologram generated by interference superposition contains a direct current (DC) component composed of object wave power and reference wave power $|E_o(\mathbf{r}_{p,s},t)|^2 + |E_r(\mathbf{r}_{p,s},t)|^2$, as well as a pair of conjugate components containing the phase difference between the two waves. Since the reference wave is generated by the radiation unit of the HIS system and its amplitude and phase information are known, we only need to calculate the amplitude and phase of the object wave using the hologram and reference wave information, after which the electromagnetic environment perception is accomplished. The specific content will be explained elaborately in Section \uppercase\expandafter{\romannumeral4}.

\section{Hologram Interference Reduction Scheme}
In this section, we explain in detail the interference composition in the hologram and how this interference affects the wave field formation. The corresponding interference reduction method is proposed based on multiple holograms. Note that in this section, we consider the $(p,s)$-th unit of the HIS and drop the arguments $(\mathbf{r}_{p,s},t)$ for ease of notation. 

\subsection{Interference Composition}
\begin{figure}[!t]
\centering
\includegraphics[width=3in]{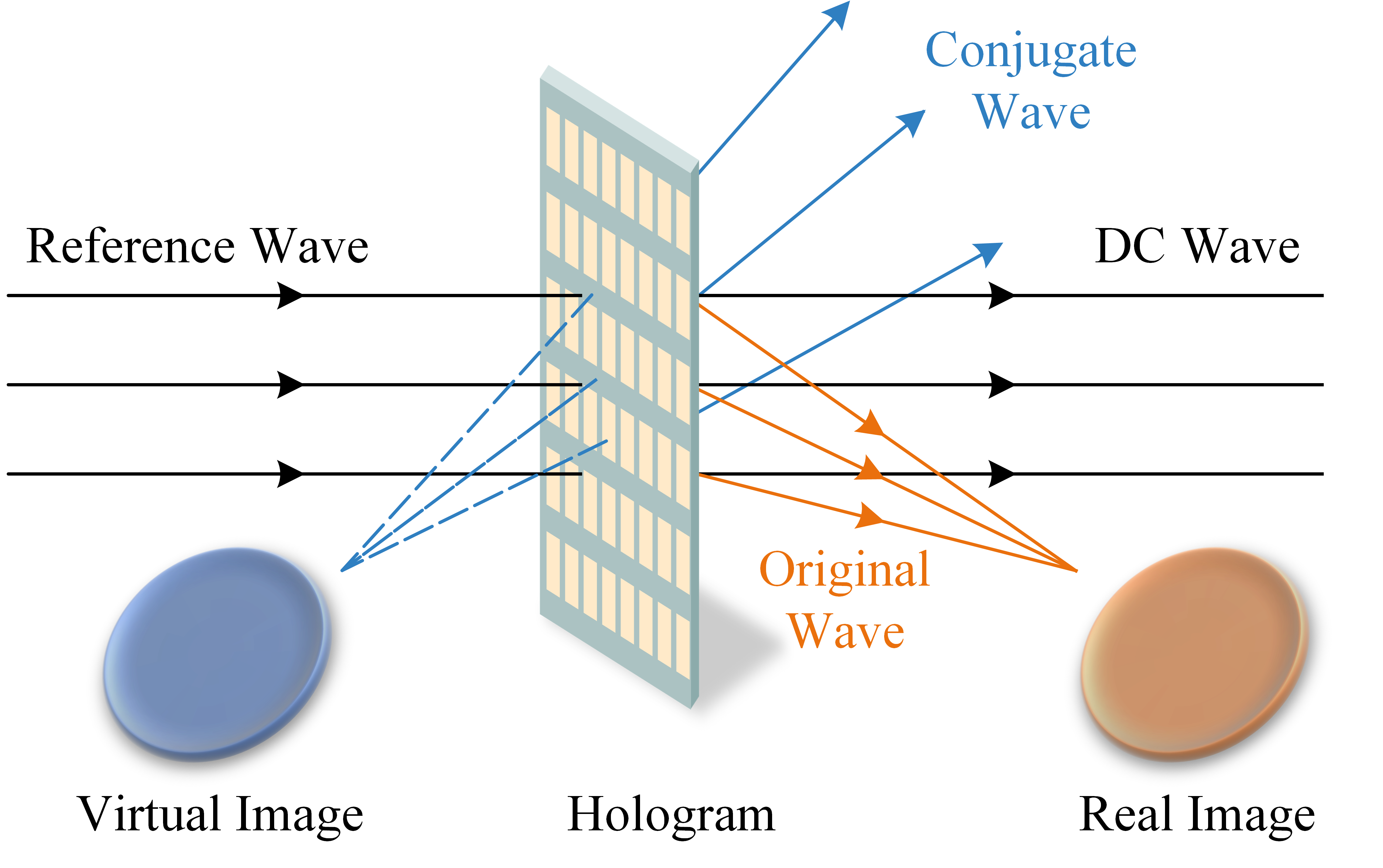}
\caption{Wave reconstruction without interference reduction.}
\label{fig_2}
\end{figure}

According to Eq. (\ref{eq15}), the hologram contains a DC component and a pair of conjugate components. If we directly take the original hologram as the emission coefficient of the antenna array for wave field formation, namely using the same reference wave to irradiate the hologram as shown in Fig. \ref{fig_2}, then the wave field in the plane where the HIS system is located is expressed as:
\begin{equation}
\label{eq16}
E_{c} = E_{r} \cdot I = (A_{r}^2 + A_{o}^2)E_{r} + A_{r}^2E_{o} + A_{r}^2e^{j2 \mathbf{k}_{r} \cdot \mathbf{r}_{p,s}}E_{o}^*.
\end{equation}

It can be seen from Eq. (\ref{eq16}) that the wave field consists of an amplitude-modulated reference wave, the original object wave, and a phase-modulated conjugate object wave. If the reference wave is uniform and its amplitude can be considered constant in the HIS, then the second term in Eq. (\ref{eq16}) is an amplitude-modulated wave whose propagation will follow the direction of the original object wave $E_o$. The third term in Eq. (\ref{eq16}) contains the conjugate form of the object wave and a phase factor. The conjugate term has similar beam characteristics with the original object wave, while its propagation direction is determined by the phase factor. It is obvious that the original hologram can not achieve the desired beamforming, and the DC and conjugate components will become the source of pollution in the wave field.

\subsection{Phase Shifting-based Interference Suppression Method}
Though communication holography and optical holography share similar fundamental principles, there are still essential differences between optic and communication with respect to signal processing requirements. First of all, different from stable optical scenes, communication scenarios have a more complex spectral composition and more variable random perturbations, requiring consideration of noise, Doppler effects, and multi-scale fading. Secondly, the communication interference suppression algorithm requires enhanced real-time performance to cope with the rapid changes in the channel. Finally, the size of communication holograms is usually much lower than optical due to the limit on the number of radiation units, which means that the spatial information is almost unavailable for communication hologram interference suppression.

The optical hologram interference suppression methods are briefly classified into five general groups: phase shifting interferometry (PSI) \cite{ref17}, off-axis techniques \cite{ref18}, \cite{ref19}, linear filtering \cite{ref20}, phase retrieval method \cite{ref21} and filtering of the complex wavefront in reconstruction planes \cite{ref22}, \cite{ref23}. Except for PSI, other methods either rely on the spatial information of holograms or require accurate CSI, which are not suitable for communication scenarios. PSI can suppress both the DC and conjugate components with a few time sampling sequences of hologram, which fits well with the characteristics of holographic communication scenarios. Therefore, we illustrate the communication hologram interference reduction method based on the PSI technique in the following.

It is assumed that the carrier frequency of users is the same as the reference wave frequency and the sample interval is always an integer multiple of the reference wave period. Without considering the movement of users, the state of the object wave emitted by users will remain stable in each sample. The reference wave only performs $\pi/2$ phase rotation before each sample, and three holograms with different phases are obtained after three samples:
\begin{align}
& I(0) = |E_{o}+E_{r}|^2 =  |E_{o}|^2 + |E_{r}|^2 + E_{o}E_{r}^* + E_{o}^*E_{r},\\
& I(\frac{\pi}{2}) = |E_{o}+jE_{r}|^2 = |E_{o}|^2 + |E_{r}|^2 - jE_{o}E_{r}^* + jE_{o}^* E_{r},\\
& I(\pi) = |E_{o}-E_{r}|^2 = |E_{o}|^2 + |E_{r}|^2 - E_{o}E_{r}^* - E_{o}^*E_{r}.
\end{align}
The conjugate component is removed based on the complex sum of two holograms:
\begin{align}
    I(0) + j\cdot I(\frac{\pi}{2}) &= (1 + j)(A_r^2 + A_o^2) + 2 E_{o}E_{r}^*,\\
    I(\frac{\pi}{2}) + j\cdot I(\pi) &= (1 + j)(A_r^2 + A_o^2) - 2 E_{o}E_{r}^*.
\end{align}
The object wave is expressed with the three holograms as:
\begin{equation}
\label{eq20}
\hat{E}_{o} = \frac{1-j}{4E_{r}^*} \left\{I(0) - I(\frac{\pi}{2}) + j\left[I(\frac{\pi}{2}) - I(\pi)\right]\right\}.
\end{equation}
Note that there are alternative strategies for choosing the phase shift step width of the reference wave and the number of hologram samples in \cite{ref24}. Assuming that the channel remains stationary, the object wave can be completely recovered through several holograms. If the carrier frequency is deviated by the Doppler effect which results in a perturbation of the object wave at each sample, it is difficult to obtain an accurate estimate of the object wave based on the above method. Therefore, there is still a need for further research to solve the problem of hologram interference reduction in mobile scenarios.

In fact, for single-user setting, the reconstructed objective wave in Eq. (\ref{eq20}) is also the CSI for the $({p,s})$-th radiation unit even in multi-path environments. The channel estimation approach for multi-user scenarios is shown in Section \uppercase\expandafter{\romannumeral5}. In traditional multiple antenna systems, the channel estimation generally requires RF chains and baseband signal processing. This is costly especially when the number of antennas is large. Compared to conventional systems, the channel estimation in our proposed architecture is implemented in radio frequency. More specifically, the channel is directly estimated with RF holograms on the surface which contains a large number of antenna elements. Therefore, our system significantly reduces the need for RF chains and circumvents the complicated processes like filtering, ADC and down conversion, which provides a new paradigm for building a more cost-effective wireless communication architecture.

\section{Prony-based Multi-user Channel Segmentation}
HIS receives signals based on hologram whose interference composition and reduction method have been studied in Section \uppercase\expandafter{\romannumeral4.} 
After applying the PSIS method, a new channel estimation algorithm is required to reconstruct the channel in multi-user scenarios. 
In this section, we first introduce the multi-user communication scenario based on the HIS. Following that, the multi-user segmentation problem for the HIS is described. A brief review of extended Prony's method is then made, based on which the segmentation algorithm for multi-user scenarios is presented.

\subsection{Scenario Description}
Consider a communication system where the BS receives data streams sent by $N$ users.
Due to the mobility and high data transmission requirements, the base station needs to accurately estimate the CSI. The BS is equipped with the HIS to segment the information of each user through the hologram. For ease of analysis, we consider the line of sight paths.

\subsection{Multi-user Channel Segmentation}

Traditional pilot-based MIMO channel estimation approaches are not appropriate for our proposed hologram-based channel sensing architecture due to two primary concerns. Firstly, conventional methods usually have excessive demand for pilot overhead because of the large number of elements at the transmit and receive ends. In addition, it is necessary to utilize at least three holograms to recover a pilot in our proposed architecture, which further increases the requirement of time-frequency resources for traditional estimation techniques. Secondly, the HIS has the potential to implement signal processing in radio frequency with reduced hardware complexity and energy consumption. However, the pilot-based techniques estimate the CSI in baseband, requiring complicated process like filtering, analog to digital conversion, down conversion. Therefore, the traditional pilot-based estimation techniques are inefficient and unable to fully exploit the hardware advantages of HIS. To cope with the above issues, we propose a multi-user channel segmentation algorithm based on extended Prony's method.

According to Eq. (\ref{eq5}), the channel between all base station units and the $u$-th UE for certain carrier frequency $f$ is simplified as:
\begin{equation}
\label{eq21}
    \mathbf{h}_u(f,t)=\beta_u e^{j\frac{2\pi\hat{r}_{rx,u}^T\overline{d}_{rx,u}}{\lambda_0}}e^{-j2\pi f_u\tau_u}e^{j\omega_ut}\mathbf{a}(\theta_u,\phi_u),
\end{equation}
where $\mathbf{a}(\theta_u,\phi_u)$ is a 3-D steering vector that is highly correlated with the physical structure of the HIS. The path loss, delay and Doppler are assumed to be constant for different HIS units within a coherence time. 

The signal transmitted by the $u$-th UE is denoted by $x_u$. According to Eq. (\ref{eq21}), the signal arriving at the $(m,n)$-th unit of the HIS after passing through the channel at time $t$ is written as:
\begin{equation}
\label{eq22}
    y(m,n)=\sum_{u=1}^N b_ue^{j(2\pi\frac{\cos{\theta_u}}{\lambda_0}D_v m+2\pi\frac{\sin{\theta_u}\sin{\phi_u}}{\lambda_0}D_h n)},
\end{equation}
where
\begin{equation}
    b_u = x_u\beta_u e^{j\frac{2\pi\hat{r}_{rx,u}^T\overline{d}_{rx,u}}{\lambda_0}}e^{-j2\pi f_u\tau_u}e^{j\omega_ut}.
\end{equation}

Noting that the HIS has symmetric nature, we take the first column units of the surface as an example to illustrate the multi-user segmentation problem. According to Eq. (\ref{eq22}), we may further write
\begin{equation}
\label{eq24}
    y(k)=\sum_{u=1}^N b_ue^{j2\pi\frac{\cos{\theta_u}}{\lambda_0}D_v k}=\sum_{u=1}^N b_u z_u^k,
\end{equation}
where
\begin{equation}
    z_u = e^{j2\pi\frac{\cos{\theta_u}}{\lambda_0}D_v}.
\end{equation}
$y(k)$ ($k=0,1,\cdots,N_v-1$) is a uniformly spatial sampled signal composed of $N$ undamped complex exponentials. The multi-user segmentation problem is to reconstruct the channel of each user by computing the message sequence $\mathbf{b}=\begin{bmatrix} b_1&\cdots&b_N\end{bmatrix}^T$ and $\mathbf{z}=\begin{bmatrix} z_1&\cdots&z_N\end{bmatrix}^T$ from the $N_v$ samples of $y(k)$, which is a typical linear parameter estimation problem.

\subsection{A Review of Extended Prony's Method}
Prony's method proposed by Gaspard Riche de Prony in 1795 is a useful tool to analyze a uniformly sampled signal composed of damped/undamped complex exponentials \cite{ref25}. The traditional Prony method is not a spectral estimation technique in the usual sense. However, with an extension, the Prony method can be used to estimate the spectral density of rational forms. A review of extended Prony method is briefly given below. Suppose we have $K$ samples of data $y(k)$ which consist of $N$ exponentially damped/undamped signals:
\begin{equation}
\label{eq25}
    y(k)=\sum_{n=1}^N b_n z_n^k,\quad 0 \leq k \leq K-1,
\end{equation}
where $b_n$ is the complex amplitude and $z_n = e^{s_n} = e^{\alpha_n+j\omega_n}$. In the special case of channel estimation, $y(k)$ can be regarded as the uniformly sampled channel estimate. Define the following polynomial:
\begin{equation}
    P_0(z)=\prod_{n=1}^N(z-z_n)=\sum_{n=0}^Np_n z^{N-n},\quad z\in\mathbb{C},
\end{equation}
where $p_0=1$ and $z_n$, ($n=1,\cdots,N$) are zeros of $P_0(z)$. For an arbitrary $m\in\mathbb{N}$, ones has
\begin{multline}
\label{eq27}
    \sum_{n=0}^N p_ny(m-n)=\sum_{n=0}^N p_n\sum_{l=1}^N b_l z_l^{m-n}\\
                          =\sum_{l=1}^N b_l\sum_{n=0}^N p_n z_l^{m-n}=\sum_{l=1}^N b_lz_l^{m-N}\sum_{n=0}^N p_n z_l^{N-n}\overset{\text{a}}{=}0,
\end{multline}
where $\overset{\text{a}}{=}$ is due to the fact that $z_l$ ($l=1,\cdots,N$) are zeros of $P_0(z)$. Eq. (\ref{eq27}) implies that the following homogeneous linear difference equation is fulfilled:
\begin{equation}
\label{eq28}
    y(k)=-\sum_{n=1}^N p_n y(k-n),\quad 0 \leq k \leq K-1.
\end{equation}

The above part is the major content of Prony's method, which provides a way to predict the tendency of signal $y$ based on Eq. (\ref{eq28}). There is an extension of Prony's method for the estimation of parameters $b_n$ and $z_n$ as below \cite{ref5}. According to Eq. (\ref{eq28}), we may obtain the coefficients $p_n$ with $K$ sampled data by solving the following linear equations:
\begin{equation}
\label{eq29}
    \mathbf{Y}\mathbf{p}=0,
\end{equation}
where $\mathbf{Y}$ is a Toeplitz matrix
\begin{align}
    \mathbf{Y}&=\begin{bmatrix}
        y(N)    &y(N-1)  &\cdots &y(0)\\
        y(N+1)  &y(N)    &\cdots &y(1)\\
        \vdots  &\vdots  &\vdots &\vdots\\
        y(K-1)  &y(K-2)  &\cdots &y(K-N-1)
    \end{bmatrix},\\
    \mathbf{p}&=\begin{bmatrix}p_0&p_1&\cdots&p_N\end{bmatrix}^T.
\end{align}

Eq. (\ref{eq29}) can be solved in the least squares sense to give the coefficients $p_n$ ($n=1,\cdots, N$) of the characteristic polynomial $P_0(z)$ whose roots are the poles $z_n$ ($n=1,\cdots,N$). The complex amplitude $b_n$ ($n=1,\cdots,N$) is obtained by solving the following equations:
\begin{equation}
\label{eq33}
    \mathbf{\Phi}\mathbf{b}=\mathbf{y},
\end{equation}
where $\mathbf{\Phi}$ is a Vandermonde matrix
\begin{align}
\label{eq34}
    \mathbf{\Phi}&=\begin{bmatrix}
        1        &1          &\cdots &1\\
        z_1      &z_2        &\cdots &z_N\\
        \vdots   &\vdots     &\vdots &\vdots\\
        z_1^{K-1}&z_2^{K-1}  &\cdots &z_N^{K-1}
    \end{bmatrix},\\
    \mathbf{b}&=\begin{bmatrix}b_1&b_2&\cdots&b_N\end{bmatrix}^T,\\
\label{eq36}
    \mathbf{y}&=\begin{bmatrix}y(0)&y(1)&\cdots&y(K-1)\end{bmatrix}^T.
\end{align}
The least squares solution to Eq. (\ref{eq33}) is given by
\begin{equation}
\label{eq37}
    \hat{\mathbf{b}} = (\mathbf{\Phi}^H\mathbf{\Phi})^{-1}\mathbf{\Phi}^H\mathbf{y}.
\end{equation}
Note that at least $2N$ samples are needed to obtain all the coefficients $p_n$ ($n=1,\cdots, N$).

\subsection{Channel Segmentation based on Extended Prony's Method}
The extended Prony's method in Section \uppercase\expandafter{\romannumeral5}. C provides a fundamental approach for linear parameter estimation which is the basic property of multi-user channel segmentation. In order to solve this problem, we propose a Prony-based multi-user channel segmentation (PMCS) method. The main idea is to estimate the signal departure angles $\theta$ and $\phi$ respectively by utilizing only the spatial samples. In the extended Prony's method, the estimation order of the characteristic polynomial $P_0(z)$ is limited to the same as the summation upper bound $N$, which means that the roots of $P_0(z)$ consist only of the parameter $z_n$ ($n=1,\cdots, N$). We can further relax the restriction on the estimation order so that the estimation order is greater than the summation upper bound $N$, which can improve the estimation accuracy of the parameter $z_n$ in the low SNR case. Suppose the signal $y(k)$ as described by Eq. (\ref{eq25}) has $K$ spatial samples. Define the following polynomial:
\begin{equation}
\label{eq38}
    H(z)=\sum_{n=0}^{P_e}h_n z^{-n},\quad h_0=1,
\end{equation}
where $P_e$ is the estimation order and $h_n$ is the coefficient of polynomial $H(z)$.

We restrict the value of $P_e$ to satisfy $N\leq P_e\leq K-N$ and obtain the coefficients $h_n$ by solving the following forward linear estimation equations:
\begin{equation}
\label{eq39}
    \mathbf{Y}\mathbf{h}=0,
\end{equation}
where
\begin{align}
\label{eq40}
    \mathbf{Y}&=\begin{bmatrix}
        y(P_e)  &y(P_e-1) &\cdots &y(0)\\
        y(P_e+1)    &y(P_e) &\cdots &y(1)\\
        \vdots  &\vdots &\vdots &\vdots\\
        y(K-1)  &y(K-2) &\cdots &y(K-P_e-1)
    \end{bmatrix},\\
    \mathbf{h}&=\begin{bmatrix}1&h_1&\cdots&h_{P_e}\end{bmatrix}^T.
\end{align}

As Kumaresan stated in \cite{ref26}, any row of $\mathbf{Y}$ can be written as a linear combination of $M$ linearly independent vectors $\mathbf{s}_n=\begin{bmatrix}1&e^{-s_n}&\cdots&e^{-P_e s_n}\end{bmatrix}$, ($n=1,\cdots,N$), $i.e.$,
\begin{equation}
    \mathbf{y}_k = \sum_{n=1}^{N} e^{(P_e+k-1)s_n}b_n \mathbf{s}_n,\quad k = 1,2,\cdots,K-P_e,
\end{equation}
where
\begin{equation}
    \mathbf{y}_k=\begin{bmatrix}
        y(P_e+k-1)&y(P_e+k-2)&\cdots&y(k-1)
    \end{bmatrix}.
\end{equation}
Since $N\leq P_e\leq K-N$, the rank of $\mathbf{Y}$ is $N$, which implies that the null space dimension of $\mathbf{Y}$ is $P_e+1-N$. As $\mathbf{h}$ is the solution vector of $\mathbf{Y}$, it follows that $\mathbf{s}_n\mathbf{h}=0$, $n=1,\cdots,N$, which indicates $H(z)$ has zeros at $z_n$ ($n=1,\cdots,N$).

Apparently, $H(z)$ has $N$ signal zeros and $P_e-N$ extraneous zeros when the restriction $N\leq P_e\leq K-N$ is fulfilled. According to \cite{ref5}, we can prove that the $P_e-N$ extraneous zeros are inside the unit circle and the signal zeros $z_n$ that satisfy $\alpha_n\geq0$ lie on or outside the unit circle.

Based on the above properties, when the poles $z_n$ satisfy the condition $\alpha_n\geq0$, the following strategy can be used to calculate the poles $z_n$ and complex amplitude $b_n$ from K samples signal $y(k)$: 1) Construct data matrix $\mathbf{Y}$ with estimation order $P_e$ satisfying $N\leq P_e\leq K-N$; 2) Obtain the minimum-norm solutions $\mathbf{h}$ by solving Eq. (\ref{eq39}) and construct characteristic polynomial $H(z)$; 3) Collect the roots of $H(z)$ on or outside the unit circle; 4) Use least square method to compute complex amplitude $b_n$ by solving Eq. (\ref{eq33}).

We now apply this method to multi-user channel segmentation. Our target is to estimate the complex amplitude $\beta_u$, delay $\tau_u$, Doppler $\omega_u$, elevation departure angle $\theta_u$ and azimuth departure angle $\phi_u$ for each user based on the $N_t$ samples $y(m,n)$ defined in Eq. (\ref{eq22}). Considering that the rows and columns units of HIS are independent, we can use the first row and first column units to calculate $\theta_u$ and $\phi_u$ separately. The samples of the first row units $y_h(k)$ and the first column units $y_v(k)$ are given by:
\begin{align}
    &y_h(k) = \sum_{u=1}^N b_{u,h}z_{u,h}^k,\quad 0\leq k\leq N_h,\\
    &y_v(k) = \sum_{u=1}^N b_{u,v}z_{u,v}^k,\quad 0\leq k\leq N_v,
\end{align}
where $b_{u,v}=b_{u,h}=b_{u} = \beta_u e^{j\frac{2\pi\hat{r}_{rx,u}^T\overline{d}_{rx,u}}{\lambda_0}}e^{-j2\pi f_u\tau_u}e^{j\omega_ut}$, $z_{u,h}=e^{j2\pi\frac{\sin{\theta_u}\sin{\phi_u}}{\lambda_0}D_h k}$ and $z_{u,v}=e^{j2\pi\frac{\cos{\theta_u}}{\lambda_0}D_v k}$. 

We need to utilize the received signals $y_h(k)$ and $y_v(k)$ to acquire the estimation of $b_{u,h}$, $b_{u,v}$, $z_{u,h}$ and $z_{u,v}$. In the following, we illustrate the process of our proposed channel segmentation algorithm based on the row sample signal $y_h(k)$.

We first indicate how to estimate $z_{u,h}$ from $y_h(k)$. According to the extended Prony's method, $z_{u,h}$ is the zero of the polynomial $H_h(z)$ on the unit circle of the complex plane, and $H_h(z)$ is given by Eq. (\ref{eq38}). Therefore, we first construct the data matrix $\mathbf{Y}_h$ using $y_h(k)$ based on Eq. (\ref{eq40}). Then solve the linear equations $\mathbf{Y}_h\mathbf{h}_h=0$ by Pseudoinverse or Total Lease Squares (TLS) \cite{ref13} to find the minimum norm solution vector $\mathbf{h}_h$ which is the coefficient of $H_h(z)$. Finally, the zeros of $H_h(z)$ on the unit circle of the complex plane are chosen as the estimation of $z_{u,h}$ denoted by $\hat{z}_{u,h}$.

The steps for estimating $b_{u,h}$ are described below. After obtaining the estimation $\hat{z}_{u,h}$, the sample sequence $y_h(k)$ constitutes a linear equation system with $b_{u,h}$ as unknown variables. Consequently, the estimation of $b_{u,h}$ is solved by linear equations $\mathbf{\Phi}_h\mathbf{b}_h=\mathbf{y}_h$, where $\mathbf{\Phi}_h$ and $\mathbf{y}_h$ are given by Eq. (\ref{eq34}) and Eq. (\ref{eq36}) respectively.

$b_{u,v}$ and $z_{u,v}$ are estimated in similar steps using column sample signal $y_v(k)$. Reconstructing the channel for a particular user requires the knowledge of $b_u$, $z_{u,v}$ and $z_{u,h}$, which reveals the association between $z_{u,v}$ and $z_{u,h}$. However, such correspondence is lost between $\hat{b}_{u,v}$ and $\hat{b}_{u,h}$ acquired by the extended Prony's method. Therefore, it is necessary to pair $\hat{b}_{u,v}$ and $\hat{b}_{u,h}$. Assuming that the parameter $b_u$ takes different values among different users, i.e:
\begin{equation}
\label{eq46}
    b_u \neq b_v,~\forall u \neq v,~u,v \in \{1,\cdots,N\}.
\end{equation}
Then we can pair $\hat{b}_{u,v}$ and $\hat{b}_{u,h}$ according to Eq. (\ref{eq46}) and the equality $b_{u,v}=b_{u,h}=b_{u}$. The whole multi-user segmentation algorithm is summarized below.

\begin{algorithm}[H]
\caption{Prony-based Multi-user Channel Segmentation.}
\begin{algorithmic}[1]
\STATE Recover $y_h(k)$ and $y_v(k)$ from the hologram based on Eq. (\ref{eq20});
\STATE Construct $\mathbf{Y}_h$ and $\mathbf{Y}_v$ respectively;
\STATE Find the minimum-norm solution vector $\mathbf{h}_h$ and $\mathbf{h}_v$;
\STATE Construct $H_h(z)$ and $H_v(z)$ and collect the zeros on the unit circle to obtain the estimation $\hat{z}_{u,h}$ and $\hat{z}_{u,v}$;
\STATE Construct $\mathbf{\Phi}_h$ and $\mathbf{\Phi}_v$ based on Eq. (\ref{eq34}) and compute the least squares estimation $\hat{b}_{u,h}$ and $\hat{b}_{u,v}$;
\STATE Pair $\hat{z}_{u,h}$ and $\hat{z}_{u,v}$ according to the value of $\hat{b}_{u,h}$ and $\hat{b}_{u,v}$.
\FOR   {$u=1,\cdots,N$}
\STATE Compute $\hat{\mathbf{h}}_u(f, t)$ based on Eq. (\ref{eq21});
\ENDFOR
\end{algorithmic}
\end{algorithm}

In the following, we summarize the complete procedure for channel estimation with HIS. When the RF signals transmitted by the UE arrive at the surface, the system generates the reference wave to interferometrically superimpose with the RF signal. The intensity of the superimposed signal is recorded as holograms by the envelope detectors. The self-interference components in the hologram is then suppressed by the PSIS method and the PMCS algorithm is applied to the processed hologram for estimating the CSI in multi-user scenarios.

Note that the PMCS algorithm estimates the CSI based on the spatial samples only, although it is unable to distinguish different users. In order to identify users, the channel estimation for each user can be implemented in a time-division approach based on the PSIS algorithm. Considering the following correlation
\begin{equation}
    N \leq P_e \leq K-N,
\end{equation}
the number of users $N$ is restricted by the size of the HIS. Since the PMCS algorithm uses singular value decomposition for noise suppression of the data matrix $\mathbf{Y}$, the size of the HIS also affects the estimation performance of the PMCS algorithm in the presence of noise, which will be analyzed in Section \uppercase\expandafter{\romannumeral5}. E. 
The complexity of our PMCS algorithm is now analyzed. The signal recovery in step 1 has a complexity order of $\mathcal{O}(N_v N_h)$. The Prony coefficient computation in step 3 has a complexity order of $\mathcal{O}(2P_e^{2.37})$ due to the matrix inversion. The complexity order of step 4 is $\mathcal{O}(P_e^{2.37})$ due to the computation of polynomial zeros. The computation of $z$ in step 5 has a complexity order of $\mathcal{O}(N^{2.37})$. The paring in step 6 has a complexity order of $\mathcal{O}(N)$. The complexity of step 7 - step 9 is $\mathcal{O}(N_v N_h N)$. Therefore, the complexity order of the PMCS algorithm is $\mathcal{O}(N_v N_h N) + \mathcal{O}(N_v N_h) + \mathcal{O}(N) + \mathcal{O}(N^{2.37}) + \mathcal{O}(3P_e^{2.37})$.

\subsection{Performance Analysis of the Segmentation Algorithm}
The asymptotical performance of our PMCS method is now analyzed. We assume that the time of interferometric recording is short enough, during which the channel is stationary. 

Before stating our main result in Theorem 1, we introduce three intermediate lemmas. Since the multi-user channel segmentation algorithm essentially aims to solve a linear estimation problem, for notational simplicity, we represent the basic problem of multi-user segmentation as Eq. (\ref{eq25}), namely $y(k)=\sum_{n=1}^N b_nz_n^k$. In practice, the received signal $\tilde{y}(k)=y(k)+w(k)$ is perturbed by hologram noise $w(k)$. Here we assume that $|w(k)|<\epsilon$ with certain accuracy $\epsilon > 0$.

\textbf{Lemma 1:}\emph{ For any $n~(n=1,\cdots,N)$, the estimation error of the PMCS algorithm satisfies}
\begin{multline}
    |\hat{z}_n - z_n| \leq\\
    \frac{\sqrt{P_e}}{|\beta_0|}\left[\frac{\|\mathbf{Y}_1^H\|_2\sqrt{K-N}\epsilon}{\sigma_1^2}+\|\mathbf{y}_0\|_2(K-N)\epsilon+\Delta \sigma_1^2\|\mathbf{h}\|_2\right].
\end{multline}

\emph{Proof:} The proof can be found in Appendix A.$\hfill\square$

\emph{Remarks:} Lemma 1 indicates that the estimation error of the PMCS method for parameter $z_n$ is mainly determined by the perturbation of the noise and the number of samples. As the number of samples increases, the error will tend to be determined by the perturbation $\epsilon$.

According to Eq. (\ref{eq37}), the estimation of $\mathbf{b}$ is given by:
\begin{equation}
    \hat{\mathbf{b}} = \mathbf{L}\mathbf{y},
\end{equation}
where
\begin{equation}
    \mathbf{L} = (\mathbf{\Phi}^H\mathbf{\Phi})^{-1}\mathbf{\Phi}^H.
\end{equation}
Denote the separation distance between the angular frequencies of two adjacent parameters $z_n=e^{j\omega_n}$ as $q$ and we have

\textbf{Lemma 2:}\emph{ Let $\omega_1<\omega_2<\cdots<\omega_N<2\pi$ with a separation distance}
\begin{equation}
    q>\frac{\pi\sqrt{2K}}{K},
\end{equation}
\emph{be given. The squared spectral norm of $\mathbf{L}$ is upper-bounded by}
\begin{equation}
    \|\mathbf{L}\|_2^2 \leq \frac{3}{K}.
\end{equation}

\emph{Proof:} The proof can be found in Appendix B.$\hfill\square$

\emph{Remarks:} Lemma 2 demonstrates that both the number of users that can be segmented by the algorithm and the accuracy of the identifiable arrival angle will increase with the number of spatial samples $K$.

\textbf{Lemma 3:}\emph{ Assume that $|\tilde{y}(k)-y(k)|<\epsilon ,(k=0,\cdots,K-1)$ and $|\hat{\omega_n}-\omega_n|<\delta ,(n=1,\cdots,N)$. The estimation error of $\mathbf{b}$ satisfies the following inequality:}
\begin{equation}
    \|\hat{\mathbf{b}}-\mathbf{b}\|_2\leq\frac{\sqrt{3}}{K}(K-\frac{1}{2})^{\frac{3}{2}}\sqrt{N}\delta\|\mathbf{y}\|_2+\sqrt{3}\epsilon.
\end{equation}

\emph{Proof:} The proof can be found in Appendix C.$\hfill\square$

\emph{Remarks:} The segmentation accuracy of parameter $b_n$ is influenced by both the number of spatial samples, estimation error of parameter $z_n$, and noise perturbation.

Gaussian noise exists when holographic interference is performed. The noise is modeled as independently and identically distributed (i.i.d.) with zero mean and a variance of $\sigma^2$, which is assumed finite.
Based on the three lemmas, our theoretical result on the asymptotic performance of the PMCS algorithm is shown in Theorem 1. 

\textbf{Theorem 1:}\emph{ For a given hologram noise power $\sigma^2$, the asymptotic performance of the PMCS algorithm yields:}
\begin{equation}
    \lim_{N_v,N_h\rightarrow\infty} \frac{\|\hat{\mathbf{h}}_{u}(f, t) - \mathbf{h}_{u}(f, t)\|_2^2}{\|\mathbf{h}_{u}(f, t)\|_2^2}=0.
\end{equation}

\emph{Proof:} The proof can be found in Appendix D.$\hfill\square$

\emph{Remarks:} Theorem 1 shows that the estimation accuracy of the PMCS algorithm is positively related to the number of units. In the presence of hologram noise, the estimation error of the PMCS algorithm converges to zero when the number of HIS units is large enough. 

\section{Numerical Results}
In this section, we evaluate the performance of the proposed multi-user channel sensing algorithm for HIS. The HIS composed of 16 rows and 16 columns of radiation units is instrumented in BS. The center carrier frequency is 3.5 GHz. 

\begin{figure}[!t]
\centering
\includegraphics[width=3.5in]{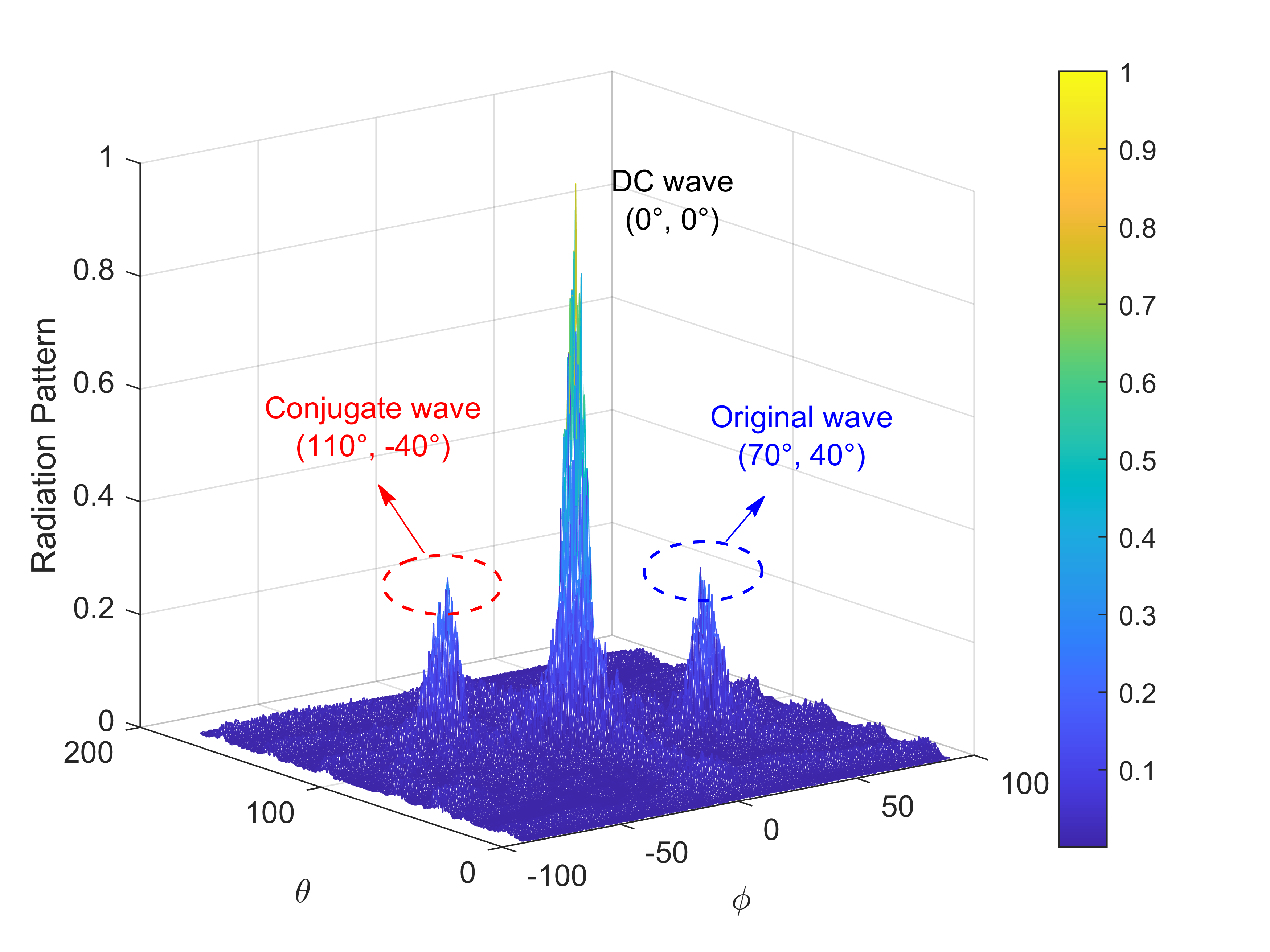}
\caption{The radiation pattern of HIS based on the original hologram.}
\label{fig_3}
\end{figure}

Fig. \ref{fig_3} illustrates the normalized radiation pattern of the HIS based on the original hologram. The direction of the user with respect to the BS is set as $(\theta,\phi)=(70^{\circ},40^{\circ})$. We perform the interferometric recording using the electromagnetic wave emitted by the user and the reference wave generated by the BS. An original hologram is constructed by the recording, which is directly adopted as the CSI. Beamforming is then performed on HIS based on the hologram. According to Eq. (\ref{eq16}), the above approaches will result in the generation of a wavefield containing DC, conjugate and original components. As can be seen in Fig. \ref{fig_3}, a relatively powerful beam is produced by the DC component in direction $(0^{\circ},0^{\circ})$, along with two sidelobes in direction $(70^{\circ},40^{\circ})$ and $(110^{\circ},-40^{\circ})$ as original and conjugate wave respectively. This demonstrates our previous analysis on the influence of the hologram self-interference components, where the DC and conjugate components will severely contaminate the wavefield.

\begin{figure}[!t]
\centering
\includegraphics[width=3.5in]{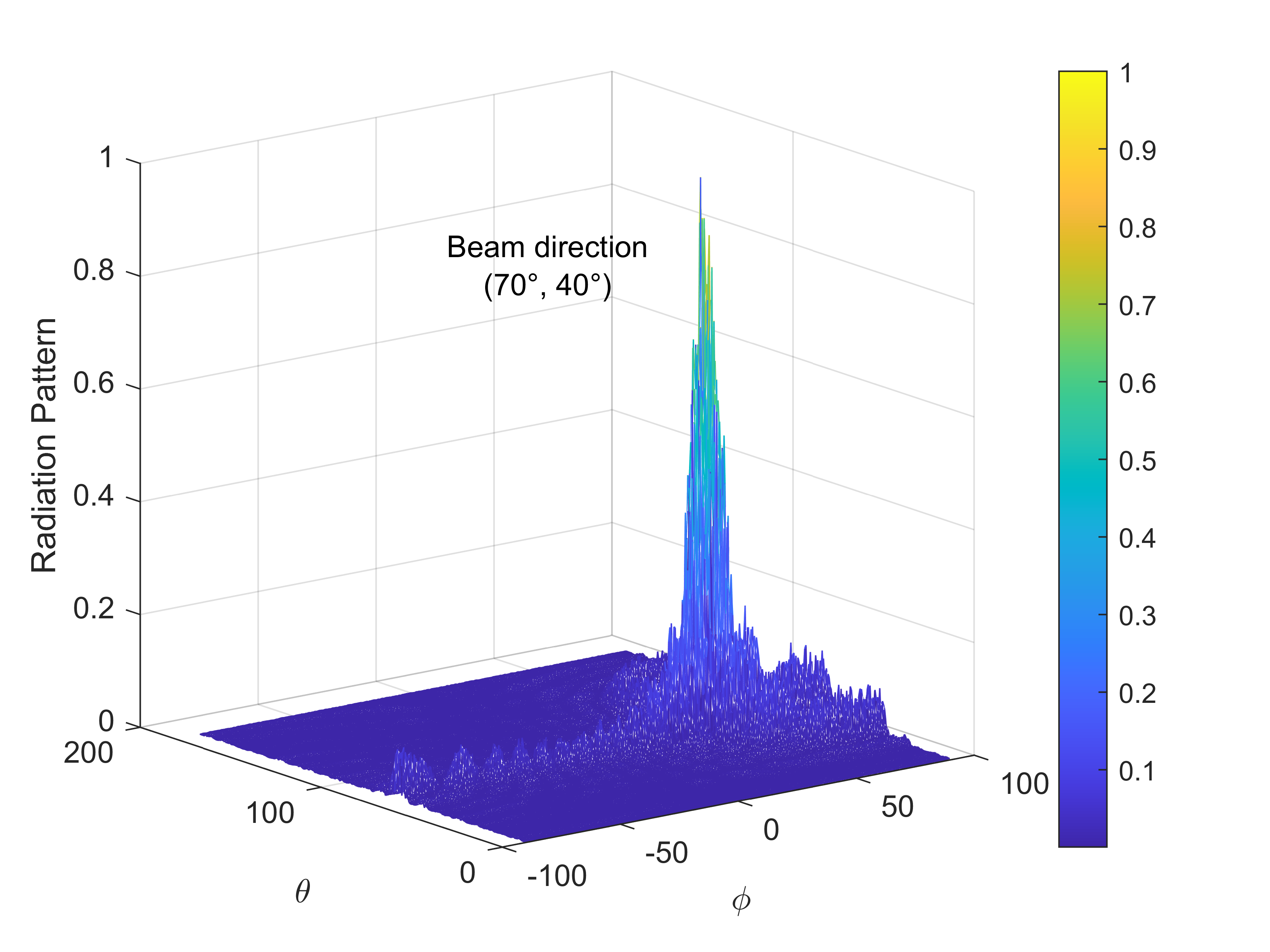}
\caption{The radiation pattern of HIS after applying PSIS method.}
\label{fig_4}
\end{figure}

Fig. \ref{fig_4} shows the HIS radiation pattern after suppressing the hologram interference. Interference suppression is performed on the hologram generated from the interferometric recording based on the PSIS method proposed in Section \uppercase\expandafter{\romannumeral4}. The processed hologram is used to estimate the CSI based on PMCS. The radiation pattern of HIS is then obtained through beamforming. It is shown in Fig. \ref{fig_4} that our proposed PSIS method effectively filters out the DC and conjugate components so as to concentrate the main energy of the array in the direction of the user.

\begin{figure}[!t]
\centering
\includegraphics[width=3.5in]{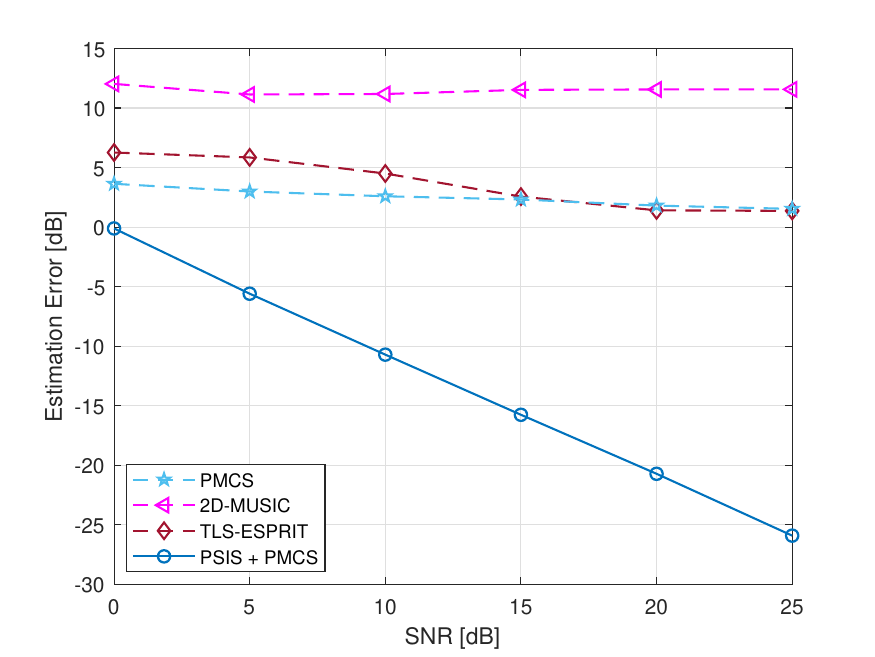}
\caption{The estimation error vs. SNR with and without the PSIS algorithm.}
\label{fig_8}
\end{figure}

Fig. \ref{fig_8} illustrates the channel estimation error with and without the PSIS algorithm. The PMCS, TLS-ESPRIT \cite{ref14}, and 2D-MUSIC \cite{ref15} algorithms are directly performed on the original hologram. The interference of the hologram is then reduced by the PSIS method and the channel estimation is accomplished by utilizing the PMCS algorithm. The estimation error is defined as
\begin{equation}
    \text{NMSE} = 10\log \left\{ \mathbb{E} \frac{\|\hat{\mathbf{h}}-\mathbf{h}\|^2_2}{\|\mathbf{h}\|^2_2} \right\}.
\end{equation}
where $\mathbf{h}$ and $\hat{\mathbf{h}}$ are the channel vector and its estimation respectively. This figure indicates that the PSIS method effectively reduces the DC and conjugate components, which significantly improves the channel estimation accuracy.

\begin{figure}[!t]
\centering
\includegraphics[width=3.5in]{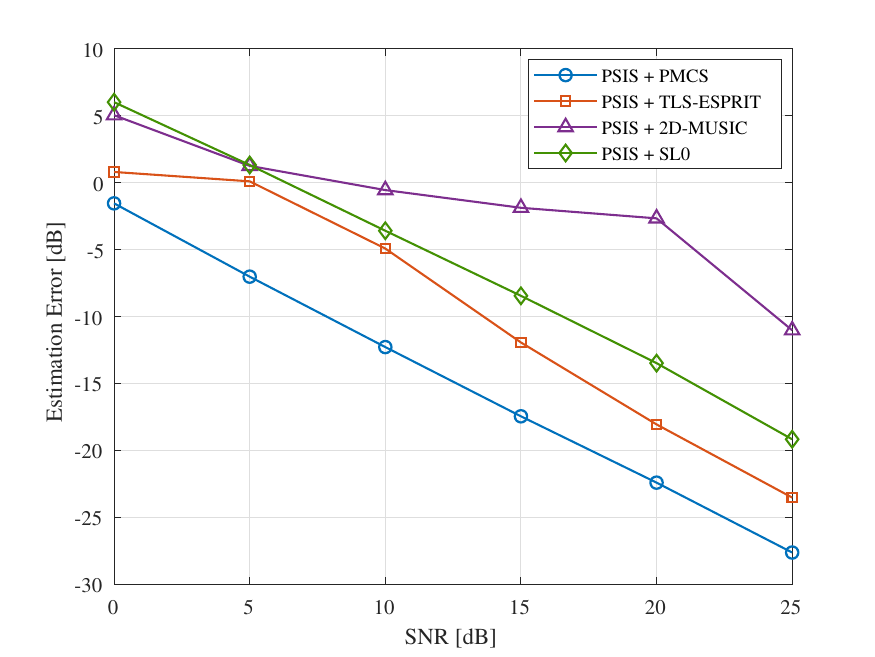}
\caption{The estimation error vs. SNR for PMCS, TLS-ESPRIT, 2D-MUSIC, and SL0 method.}
\label{fig_5}
\end{figure}

Fig. \ref{fig_5} compares the performance of our proposed PMCS algorithm with TLS-ESPRIT, 2D-MUSIC and a compressed sensing-based algorithm SL0 \cite{ref35}, \cite{ref36} in terms of the SNR and the channel estimation error. TLS-ESPRIT and 2D-MUSIC algorithms are used to estimate the angles of each user and the estimation of $\mathbf{b}$ is still accomplished based on Eq. (\ref{eq37}). The complexity order of TLS-ESPRIT, 2D-MUSIC and SL0 is $\mathcal{O}(N_vN_hN)+\mathcal{O}(N_v^{2.37})$, $\mathcal{O}(N_v^{2.37}) + \mathcal{O}(G_vG_h)$ and $\mathcal{O}(N_t^{2.37})$ respectively, where $G_v$ and $G_h$ are the angular resolution of 2D-MUSIC. The estimation order of the PMCS algorithm is set as 10. TLS-ESPRIT and 2D-MUSIC perform two time samples along with spatial samples, while the PMCS algorithm requires spatial samples only. The number of measurements for SL0 is set as 256. We may observe from Fig. \ref{fig_5} that our PMCS method, although consuming only half time resources, provides better estimation than TLS-ESPRIT, 2D-MUSIC and SL0.

\begin{figure}[!t]
\centering
\includegraphics[width=3.5in]{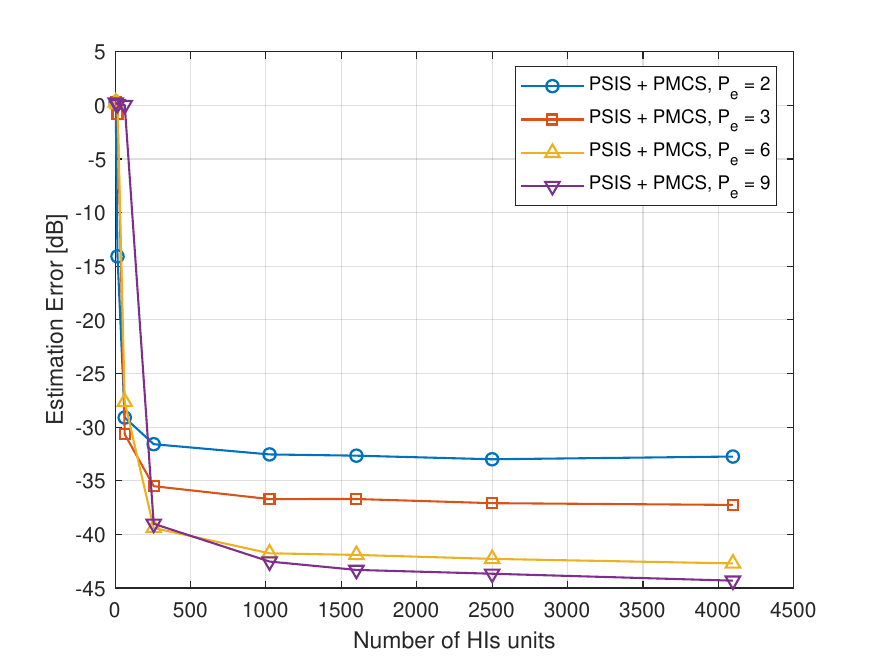}
\caption{The estimation error vs. the number of HIS units.}
\label{fig_6}
\end{figure}

Fig. \ref{fig_6} shows the channel estimation error of the PMCS method as a function of the number of HIS units, which are $N_t = 4,16,64,256,1024,1600,2500,4096$. The arrival angles of users are set as $(\theta,\phi)=(90^{\circ},0^{\circ})$ and $(30^{\circ},60^{\circ})$. This figure confirms our analysis in Theorem 1 that the estimation error of our PMCS algorithm keeps decreasing when the number of HIS units increases.

\begin{figure}[!t]
\centering
\includegraphics[width=3.5in]{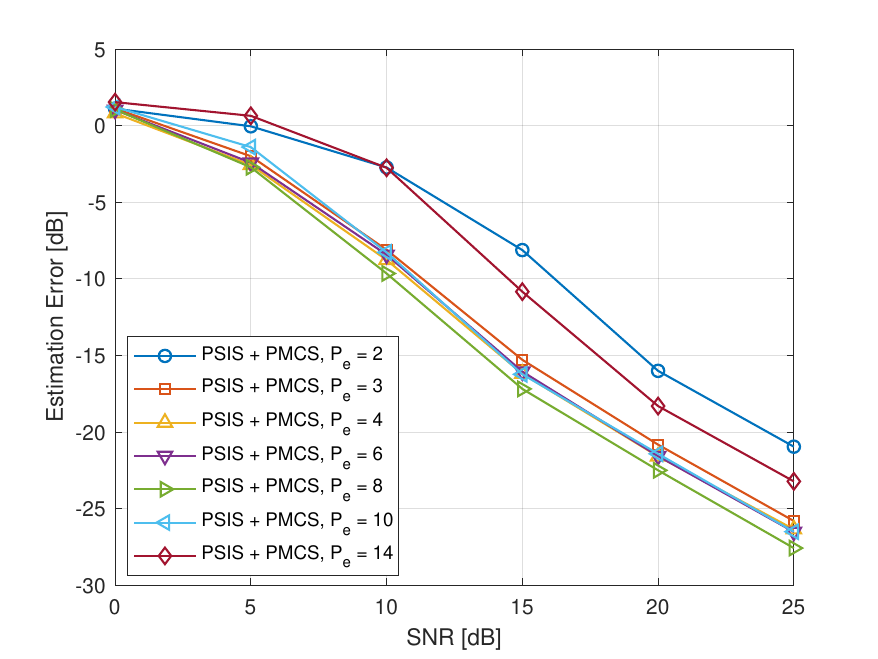}
\caption{The estimation error vs. SNR for different estimation orders.}
\label{fig_7}
\end{figure}

The estimation order $P_e$ determines the number of zeros of the estimation polynomial $H(z)$, which will affect the accuracy of the algorithm for the estimation of the parameter $z_{u,v}$ and $z_{u,h}$. We keep the number of users and arrival angles constant, adjust the value of estimation order $P_e$, and simulate under different SNRs to obtain Fig. \ref{fig_7}. From this figure, it can be seen that the effect of estimation order variation on the channel estimation error is relatively small, and the channel estimation error increases slightly when the estimation order reaches its boundary $N\leq P_e \leq K-N$. When $P_e$ reaches its lower bound, $H(z)$ will contain only the signal zeros $z_{u,v}$ or $z_{u,h}$, which means that the offset range of zeros will increase under the influence of noise. When $P_e$ reaches its upper bound, $H(z)$ will contain a large number of additional zeros, and the offset of the additional zeros will likely affect the identification and extraction of the signal zeros as the SNR decreases.

\section{Conclusion}
In this paper, we presented an interference principle-based channel sensing architecture for HIS. A PSIS method for hologram interference suppression was proposed, in order to solve the self-interference and DC problem. With the denoised hologram recording, We proposed a multi-user segmentation algorithm based on the extended Prony's method to estimate the channel for HIS. The proposed PMCS algorithm adequately considers the correlation between HIS units under UPA channel model. The channel state information of each user can be accurately estimated. Our theoretical analysis proved that the PMCS algorithm is able to achieve asymptotically error-free CSI estimation. Simulation results have also verified the effectiveness of our method which provides asymptotically error-free estimation as the number of HIS units increases.

Finally, our work opens a new prospect to signal receiving and channel estimation by only dealing with the power of the received signal. Our theoretical analysis and simulation results show that the communication architecture based on holographic interference surface provides a simplified and intuitive processing paradigm for signal receiving and channel estimation. Holographic interference surfaces break the hardware limit on the number of antennas and provide a new way to effectively deploy massive MIMO arrays in practice.

\appendix
\subsection{Proof of Lemma 1}
The parameter $z_n$ is the zeros of the polynomial $H(z)$ lying on the unit circle of the complex plane. The coefficients of the polynomial $H(z)$ can be obtained by solving Eq. (\ref{eq39}) which can be transformed into the following form
\begin{equation}
    \tilde{\mathbf{Y}}_1 \tilde{\mathbf{b}} = -\tilde{\mathbf{y}}_0,
\end{equation}
where
\begin{align}
    \tilde{\mathbf{Y}}_1 &= \begin{bmatrix}
        \tilde{y}(P_e-1) & \tilde{y}(P_e-2) & \cdots & \tilde{y}(0)\\
        \tilde{y}(P_e)   & \tilde{y}(P_e-1) & \cdots & \tilde{y}(1)\\
        \vdots           & \vdots           & \vdots & \vdots      \\
        \tilde{y}(K-1)   & \tilde{y}(K-P_e) & \cdots & \tilde{y}(K-P_e-1)\\
    \end{bmatrix},\\
    \tilde{\mathbf{y}}_0 &= \begin{bmatrix}
        \tilde{y}(P_e) & \tilde{y}(P_e+1) & \cdots & \tilde{y}(K-1)
    \end{bmatrix}^T.
\end{align}
The estimation coefficients are obtained by performing a reduced rank pseudoinverse of matrix $\tilde{\mathbf{Y}}_1$:
\begin{equation}
\label{eq58}
    \tilde{\mathbf{h}} = -\tilde{\mathbf{Y}}_1^{\dagger}\tilde{\mathbf{y}}_0,
\end{equation}
When the SNR is relatively high, the rank of matrix $\tilde{\mathbf{Y}}_1$ is approximately one, so we can further write
\begin{equation}
    \tilde{\mathbf{Y}}_1^{\dagger} = \frac{\tilde{\mathbf{Y}}_1^H}{\tilde{\sigma}_1^2},
\end{equation}
where $\tilde{\sigma}_1$ is the principal singular value of matrix $\tilde{\mathbf{Y}}_1$. According to Eq. (\ref{eq58}), we can conclude that:
\begin{equation}
    \tilde{\mathbf{h}} = -\frac{1}{\tilde{\sigma}_1^2}\tilde{\mathbf{Y}}_1^{H}\tilde{\mathbf{y}}_0,
\end{equation}
Let $\tilde{\mathbf{Y}}_1 = \mathbf{Y}_1 + \mathbf{W}_1$ and $\tilde{\mathbf{y}}_0 = \mathbf{y}_0 + \mathbf{w}_0$, and the first order perturbation in the coefficients is expressed as
\begin{equation}
    \Delta \mathbf{h} = \tilde{\mathbf{h}} - \mathbf{h} = -\frac{1}{\sigma_1^2}[\mathbf{Y}_1^H\mathbf{w}_0 + \mathbf{W}_1^H\mathbf{y}_0 + \mathbf{h}(\tilde{\sigma}_1^2 - \sigma_1^2)],
\end{equation}
According to the expression of $\Delta \mathbf{h}$, its norm is upper bounded by
\begin{align}
\label{eq62}
    \|\Delta \mathbf{h}\|_2 &= \frac{1}{\sigma_1^2}\|\mathbf{Y}_1^H\mathbf{w}_0\|_2 + \|\mathbf{W}_1^H\mathbf{y}_0\|_2 + \|\mathbf{h}(\tilde{\sigma}_1^2 - \sigma_1^2)\|_2 \\
    &\leq \frac{1}{\sigma_1^2}\|\mathbf{Y}_1^H\|_2\|\mathbf{w}_0\|_2 + \|\mathbf{W}_1^H\|_2\|\mathbf{y}_0\|_2 + (\tilde{\sigma}_1^2 - \sigma_1^2)\|\mathbf{h}\|_2
\end{align}
Since $|\tilde{y}(k)-y(k)|<\epsilon$, the spectral norm $\|\mathbf{w}_0\|_2$ and $\|\mathbf{W}_1^H\|_2$ satisfies 
\begin{align}
    \|\mathbf{w}_0\|_2   &\leq \sqrt{K-P_e}\epsilon \leq \sqrt{K-N}\epsilon, \\
    \|\mathbf{W}_1^H\|_2 &\leq \sqrt{\|\mathbf{W}_1^H\|_1\|\mathbf{W}_1^H\|_\infty} \\
                         &\leq \sqrt{(K-P_e)P_e}\epsilon \leq (K-N)\epsilon.
\end{align}
Since the perturbation $\Delta \mathbf{h}$ in the coefficients results in the deviation of the root $z_n$ of the polynomial $H(z)$, the deviation $\Delta z_n = \tilde{z}_n - z_n~(n = 1,2,\cdots,N)$ is given by
\begin{equation}
\label{eq65}
    \Delta z_n = -\frac{1}{\beta_0}\mathbf{g}_0^H\Delta \mathbf{h},
\end{equation}
where
\begin{align}
    & \mathbf{g}_0 = \begin{bmatrix}
        z_n^{P_e-1} & z_n^{P_e-2} & \cdots & 1
    \end{bmatrix}^H,\\
    & \beta_0 = \prod_{i=1}^{P_e-1} (z_n - z_i).
\end{align}
$\{z_i\}_{i=1}^{P_e-1}$ denotes the roots other than $z_n$ of the polynomial $H(z)$. Based on Eq. (\ref{eq65}), the absolute value of $\Delta z_n$ is written as
\begin{equation}
\label{eq68}
    |\Delta z_n| = \frac{1}{|\beta_0|}|\mathbf{g}_0^H\Delta \mathbf{h}| \leq \frac{1}{|\beta_0|}\|\mathbf{g}_0^H\|_2\|\Delta \mathbf{h}\|_2,
\end{equation}
Since parameter $z_n$ is distributed on the unit circle, $\|\mathbf{g}_0^H\|_2$ is given by
\begin{equation}
\label{eq69}
    \|\mathbf{g}_0^H\|_2 = \sqrt{\sum_{k=1}^{P_e} |z_n|^{2(P_e-k)}} = \sqrt{P_e}.
\end{equation}
Take Eq. (\ref{eq62}) and Eq. (\ref{eq69}) into Eq. (\ref{eq68}), we obtain
\begin{multline}
    |\Delta z_n| \leq\\
    \frac{\sqrt{P_e}}{|\beta_0|}\left[\frac{\|\mathbf{Y}_1^H\|_2\sqrt{K-N}\epsilon}{\sigma_1^2}+\|\mathbf{y}_0\|_2(K-N)\epsilon+\Delta \sigma_1^2\|\mathbf{h}\|_2\right].
\end{multline}
which proves Lemma 1.$\hfill\square$

\subsection{Proof of Lemma 2}
According to Eq. (\ref{eq34}), the matrix $\mathbf{\Phi}^H\mathbf{\Phi}$ is given by
\begin{equation}
    \mathbf{\Phi}^H\mathbf{\Phi} = \begin{bmatrix}
        \gamma_{11} & \gamma_{12} & \cdots & \gamma_{1N} \\
        \gamma_{21} & \gamma_{22} & \cdots & \gamma_{2N} \\
        \vdots      & \vdots      & \vdots & \vdots      \\
        \gamma_{N1} & \gamma_{N2} & \cdots & \gamma_{NN} \\
    \end{bmatrix},
\end{equation}
where
\begin{equation}
\label{eq72}
    \gamma_{ij}=\frac{(z_i^*z_j)^K-1}{z_i^*z_j-1},
\end{equation}
We use Gershgorin's Disk Theorem such that for an arbitrary eigenvalue $\lambda$ of the matrix $\mathbf{\Phi}^H\mathbf{\Phi}$, we preserve the inequality
\begin{equation}
\label{eq73}
    |\lambda-\gamma_{ii}|=|\lambda-K|\leq \max\{\sum_{\substack{j=1\\i\neq j}}^N |\gamma_{ij}|;i=1,2,\cdots,N\}.
\end{equation}
Based on Eq. (\ref{eq72}), $|\gamma_{ij}|$ is written in the form
\begin{align}
    |\gamma_{ij}|&=\sqrt{\frac{e^{j(\omega_j-\omega_i)N}-1}{e^{j\omega_j-\omega_i}-1}\cdot\frac{e^{-j(\omega_j-\omega_i)N}-1}{e^{-j(\omega_j-\omega_i)}-1}}\\
                 &=\sqrt{\frac{1-\cos{(N(\omega_j-\omega_i))}}{1-\cos{(\omega_j-\omega_i)}}}.
\end{align}
Apparently, we can further obtain the following inequality
\begin{align}
    |\gamma_{ij}|&=\sqrt{\frac{1-\cos{(N(\omega_j-\omega_i))}}{1-\cos{(\omega_j-\omega_i)}}}\\
                 &=\left\vert\frac{\sin{(\frac{N(\omega_j-\omega_i)}{2})}}{\sin{(\frac{\omega_j-\omega_i}{2})}}\right\vert\leq \left|\sin{(\frac{\omega_j-\omega_i}{2})}\right|^{-1}.
\end{align}
Considering that the angular frequencies $\omega_n$ is distributed in the range $[0,2\pi]$, $\frac{\omega_j-\omega_i}{2}$ is restricted to the range $[-\frac{\pi}{2},\frac{\pi}{2}]$. Therefore, using the inequality
\begin{equation}
    |\sin{x}| \geq \left|\frac{2}{\pi}x\right|,\quad x\in [-\frac{\pi}{2},\frac{\pi}{2}],
\end{equation}
we may obtain
\begin{align}
    \sum_{\substack{j=1\\i\neq j}}^N |\gamma_{ij}| \leq \sum_{\substack{j=1\\i\neq j}}^N \left|\sin{(\frac{\omega_j-\omega_i}{2})}\right|^{-1} \leq \sum_{\substack{j=1\\i\neq j}}^N \left|\frac{\omega_j-\omega_i}{\pi}\right|^{-1}.
\end{align}
Since $q$ denotes the smallest separation distance between two adjacent angular frequencies and the angular frequencies distribution is within the range $[0,2\pi]$, $q$ must satisfy the inequality $qN < 2\pi$. Then we may further write
\begin{align}
\label{eq78}
     \sum_{\substack{j=1\\i\neq j}}^N |\gamma_{ij}| \leq \sum_{\substack{j=1\\i\neq j}}^N |\frac{\omega_j-\omega_i}{\pi}|^{-1} \leq \sum_{\substack{j=1\\i\neq j}}^N |\frac{jq}{\pi}|^{-1} < \frac{N\pi}{q}.
\end{align}
Based on Eq. (\ref{eq78}) and the inequality $qN < 2\pi$, we can write the upper bound of Eq. (\ref{eq73}) as 
\begin{align}
\label{eq79}
    |\lambda-\gamma_{ii}|&=|\lambda-K|\\
                         &\leq \max\{\sum_{\substack{j=1\\i\neq j}}^N |\gamma_{ij}|;i=1,2,\cdots,N\}\\
                         &<\frac{N\pi}{q}<\frac{2\pi^2}{q^2}.
\end{align}
Let $\lambda_{min}$ and $\lambda_{max}$ be the smallest and largest eigenvalue of $\mathbf{\Phi}^H\mathbf{\Phi}$, respectively. According to Eq. (\ref{eq79}), we obtain
\begin{align}
    K-\frac{2\pi^2}{q^2}\leq \lambda_{min} \leq K \leq \lambda_{max} \leq K+\frac{2\pi^2}{q^2}.
\end{align}
Denote $\sigma_{max}$ and $\sigma_{min}$ as the largest and smallest singular value respectively. From the relationship between the singular values and the spectral norm, we can obtain
\begin{equation}
    \|\mathbf{\Phi}^H\|_2^2=\|\mathbf{\Phi}^H\mathbf{\Phi}\|_2=\sigma_{max}^2.
\end{equation}
According to the singular value property of the pseudoinverse matrix, there is also a reciprocal inverse relationship between the singular values of matrix $\mathbf{A}$ and its pseudoinverse matrix $\mathbf{A}^{\dagger}$. Therefore, we can obtain the following relationship
\begin{equation}
    \|(\mathbf{\Phi}^H\mathbf{\Phi})^{-1}\|_2=\frac{1}{\sigma_{min}^2}.
\end{equation}
Since the non-zero singular value of matrix $\mathbf{A}$ is the square root of the non-zero eigenvalue of matrix $\mathbf{A}^H\mathbf{A}$, we can obtain the following inequality
\begin{align}
    \|\mathbf{L}\|_2^2  &=\|(\mathbf{\Phi}^H\mathbf{\Phi})^{-1}\mathbf{\Phi}^H\|_2^2\\
                        &\leq \|(\mathbf{\Phi}^H\mathbf{\Phi})^{-1}\|_2^2\|\mathbf{\Phi}^H\|_2^2=\frac{\sigma_{max}^2}{\sigma_{min}^4}=\frac{\lambda_{max}}{\lambda_{min}^2}\\
                        &\leq \frac{K+\frac{2\pi^2}{q^2}}{(K-\frac{2\pi^2}{q^2})^2}<\frac{3}{K},
\end{align}
which proves Lemma 2.$\hfill\square$

\subsection{Proof of Lemma 3}
The estimate of parameter $\mathbf{b}$ is obtained by solving the overdetermined linear system described in Eq. (\ref{eq34}) using the least squares method, thus it follows that
\begin{align}
    \|\hat{\mathbf{b}}-\mathbf{b}\|_2&=\|(\mathbf{\Phi}^H\mathbf{\Phi})^{-1}\mathbf{\Phi}^H\tilde{\mathbf{y}}-\mathbf{b}\|_2\\
                                     &=\|\hat{\mathbf{L}}\tilde{\mathbf{y}}-\mathbf{b}\|_2=\|\hat{\mathbf{L}}\hat{\mathbf{Z}}\mathbf{b}-\hat{\mathbf{L}}\mathbf{Z}\mathbf{b}-\hat{\mathbf{L}}(\tilde{\mathbf{y}}-\mathbf{y})\|_2\\
                                     &\leq \|\hat{\mathbf{L}}\|_2\|\hat{\mathbf{Z}}-\mathbf{Z}\|_2\|\mathbf{b}\|_2+\|\hat{\mathbf{L}}\|_2\|\tilde{\mathbf{y}}-\mathbf{y}\|_2.
\end{align}
The squared spectral norm of $\hat{\mathbf{Z}}-\mathbf{Z}$ is upper-bounded by the Frobenius norm
\begin{align}
\label{eq86}
    \|\hat{\mathbf{Z}}-\mathbf{Z}\|_2^2 &\leq \|\hat{\mathbf{Z}}-\mathbf{Z}\|_F^2=\sum_{n=1}^N \sum_{k=0}^{K-1} |e^{j\hat{\omega}_n k}- e^{j\omega_n k}|^2\\
    &=\sum_{n=1}^N \sum_{k=0}^{K-1}[2-2\cos{(k(\hat{\omega}_n-\omega_n))}]\\
    &=\sum_{n=1}^N \left[2K-1-\frac{\sin{((K-\frac{1}{2})(\hat{\omega}_n-\omega_n))}}{\sin{(\frac{\hat{\omega}_n-\omega_n}{2})}}\right].
\end{align}
Applying the following property of the Dirichlet kernel
\begin{align}
    \frac{\sin{\frac{(2m-1)x}{2}}}{\sin{\frac{x}{2}}} \geq 2m-1+\left[-\frac{1}{3}(m-\frac{1}{2})^3+\frac{m}{12}-\frac{1}{24}\right]x^2,
\end{align}
we may further write Eq. (\ref{eq86}) as
\begin{align}
\label{eq88}
    \|\hat{\mathbf{Z}}-\mathbf{Z}\|_2^2 &\leq \sum_{n=1}^N {\left[\frac{1}{3}(K-\frac{1}{2})^3-\frac{K}{12}+\frac{1}{24}\right](\hat{\omega}_n-\omega_n)^2}\\
    &\leq \frac{1}{3}(K-\frac{1}{2})^3N\delta^2,
\end{align}
Based on Eq. (\ref{eq88}), the upper bound of $\|\hat{\mathbf{b}}-\mathbf{b}\|_2$ is written as
\begin{align}
    \|\hat{\mathbf{b}}-\mathbf{b}\|_2 \leq \sqrt{\frac{N}{3}}(K-\frac{1}{2})^{\frac{3}{2}}\delta\|\hat{\mathbf{L}}\|_2\|\mathbf{b}\|_2+\|\hat{\mathbf{L}}\|_2\|\tilde{\mathbf{y}}-\mathbf{y}\|_2.
\end{align}
The upper bound of $\|\hat{\mathbf{L}}\|_2$ is obtained from \emph{Lemma 2} which follows
\begin{equation}
    \|\hat{\mathbf{L}}\|_2 < \sqrt{\frac{3}{K}}.
\end{equation}
Finally, we use
\begin{align}
    & \|\tilde{\mathbf{y}}-\mathbf{y}\|_2 \leq \sqrt{K}\|\tilde{\mathbf{y}}-\mathbf{y}\|_{\infty}\leq \sqrt{K}\epsilon,\\
    & \|\mathbf{b}\|_2 = \|\mathbf{L}\mathbf{y}\|_2 \leq \|\mathbf{L}\|_2\|\mathbf{x}\|_2 < \sqrt{\frac{3}{K}}\|\mathbf{x}\|_2,
\end{align}
then the upper bound of $\|\hat{\mathbf{b}}-\mathbf{b}\|_2$ is expressed as
\begin{equation}
    \|\hat{\mathbf{b}}-\mathbf{b}\|\leq\frac{\sqrt{3}}{K}(K-\frac{1}{2})^{\frac{3}{2}}\sqrt{N}\delta\|\mathbf{y}\|+\sqrt{3}\epsilon.
\end{equation}
which proves Lemma 3.$\hfill\square$

\subsection{Proof of Theorem 1}
According to Eq. (\ref{eq21}), the $l$-th element of the channel vector $\mathbf{h}_u(f, t)$ between UE and the base station is written as
\begin{equation}
    \mathbf{h}_{u,l}(f, t) = z_{u,h}^mz_{u,v}^nb_u,
\end{equation}
where 
\begin{align}
     z_{u,h} &= e^{j2\pi\frac{D_h\sin{\theta}\sin{\phi}}{\lambda_0}},\\
     z_{u,v} &= e^{j2\pi\frac{d_v\cos{\theta}}{\lambda_0}},\\
     b_u &= \beta_u e^{j\frac{2\pi\hat{r}_{rx,u}^T\overline{d}_{rx,u}}{\lambda_0}}e^{-j2\pi f_u\tau_u}e^{j\omega_ut}.
\end{align}
Let $\hat{z}_{u,h}=z_{u,h}+e_h$, $\hat{z}_{u,v}=z_{u,v}+e_v$ and $\hat{b}_{u}=b_{u}+e_b$. According to the inequality
\begin{equation}
    |e^{jx}-e^{jy}| = \left|\int_x^y e^{jt} dt\right| \leq \left|\int_x^y 1 dt\right| = |y - x|,
\end{equation}
the error between $\mathbf{h}_{u,l}(f, t)$ and its estimation is upper bounded by
\begin{multline}
    |\hat{\mathbf{h}}_{u,l}(t) - \mathbf{h}_{u,l}(t)| = |\hat{z}_{u,h}^m\hat{z}_{u,v}^n\hat{b}_u - z_{u,h}^m z_{u,v}^n b_{u}| \\ 
    \leq mnk|\text{phase}(\hat{z}_{u,h}\hat{z}_{u,v}\hat{b}_u)-\text{phase}(z_{u,h} z_{u,v} b_{u})|,
\end{multline}
where $k =$ max\{$|\hat{z}_{u,h}\hat{z}_{u,v}\hat{b}_u|$, $|z_{u,h} z_{u,v} b_{u}|$\}. According to \emph{Lemma 1} and \emph{Lemma 3}, the absolute values $|e_h|$, $|e_v|$ and $|e_b|$ is bounded by the variation $\epsilon$ and $\delta$, which means the error $|\hat{\mathbf{h}}_{u,l}(t) - \mathbf{h}_{u,l}(t)|$ is limited for certain $\epsilon$ and $\delta$.
Therefore,
\begin{equation}
    \lim_{N_v,N_h\rightarrow\infty} \frac{\|\hat{\mathbf{h}}_{u}(t) -\mathbf{h}_{u}(t)\|_2^2}{N_vN_h\beta_u^2}=0.
\end{equation}
Notice that
\begin{equation}
    \lim_{N_v,N_h\rightarrow\infty} \frac{\|\mathbf{h}_{u}(t)\|_2^2}{N_vN_h}=\beta_u^2,
\end{equation}
we may further derive
\begin{align}
    &\lim_{N_v,N_h\rightarrow\infty} \frac{\|\hat{\mathbf{h}}_{u}(t) - \mathbf{h}_{u}(t)\|_2^2}{\|\mathbf{h}_{u}(t)\|_2^2}=\\
    &\lim_{N_v,N_h\rightarrow\infty} \frac{\|\hat{\mathbf{h}}_{u}(t) - \mathbf{h}_{u}(t)\|_2^2}{N_vN_h\beta_u^2}=0.
\end{align}
which proves Theorem 1.$\hfill\square$

\end{document}